\documentclass[twocolumn,pra,aps,showpacs]{revtex4}
\usepackage{xspace,amsmath,amsfonts,amsthm,amssymb,amsbsy,graphicx,color}

\begin{document}

\title{Linear quantum trajectories: Applications to continuous projection measurements}
\author{K. Jacobs and P. L. Knight}
\address{Optics Section, The Blackett Laboratory, Imperial College, 
London SW7 2BZ, England}

\begin{abstract}
We present a method for obtaining evolution operators for linear quantum 
trajectories. We apply this to a number of physical examples of varying 
mathematical complexity, in which the 
quantum trajectories describe the continuous projection measurement of 
physical observables. Using this method we calculate 
the average conditional uncertainty for the measured observables, being a 
central quantity of interest in these measurement processes.
\end{abstract}

\maketitle

\section{Introduction}
Quantum master equations, which govern the evolution of a density matrix 
representing the state of a physical system, have a wide application in
quantum 
dissipation and continuous measurement theory~\cite{BHE,Arg,WandM}. They 
describe the evolution of a quantum system that is interacting with an 
environment which, due to the interaction, may absorb energy from the system 
(dissipation), and will continually provide information about the state of the 
system (continuous measurement). A classic example of a system interacting with 
an environment is that of a single mode of an optical cavity which is allowed 
to 
leak out of the cavity via an imperfect end-mirror. The photons in the cavity 
leak out over time, and these may be detected with a photo-detector. The 
environment 
consists of the continuum of optical modes outside the cavity, and provides a 
mechanism for dissipation and continuous measurement. A master equation would 
describe the evolution of the system averaged over all the possible times at 
which the photons may be detected leaving the cavity. However, the master 
equation 
may be rewritten in an equivalent form as a stochastic equation which describes 
the evolution of the system for {\em each} set of photo-detection 
times~\cite{nonLSE}. Each possible realisation of the stochastic equation 
corresponds to a set of detection times, or more generally, to a particular 
set of measurement results. Each set of results is termed a quantum 
{\em trajectory}~\cite{Hcarm}, and the stochastic equation is said to {\em 
unravel} 
the master equation. The kind of stochastic process appearing in the equation 
will depend upon the kind of measurement process. For photo-detection of the 
output of a cavity mode the stochastic process is a point process~\cite{Point}, 
while for homodyne detection it is a Wiener process~\cite{StochMeth}. However, 
the master equation, giving the overall average evolution, does not depend upon 
the choice of measurement. In other words, there are many different ways to 
unravel any particular master equation. Here we will be concerned with 
stochastic equations which contain the Wiener process.

The fact that quantum master equations may be rewritten as Linear Stochastic 
Equations (LSE's) for the quantum state vector, referred to alternatively as 
{\em linear quantum trajectories}, has been known in the mathematical physics 
literature for some time~\cite{mathLSE}, but has only fairly recently seen 
exposure in the physics literature~\cite{GandG,HMWetc,Strunz}, where it has 
been common to use non-linear stochastic equations~\cite{solNLSE,BS}. The 
advantage of writing master 
equations as LSE's, rather than the more familiar non-linear version, is that 
in certain cases it has been found that explicit evolution operators 
corresponding to these equations may be obtained in a straightforward manner. 
However, as far as we are aware, the only method that has been used to obtain 
evolution operators for these equations to date is to choose an initial state 
which allows the stochastic equation for the state to be written as 
a stochastic 
equation for an eigenvalue, or which simplifies the action of the evolution 
operator~\cite{GandG,HMWetc}. In this paper we present a more general method 
for obtaining explicit evolution operators for 
these equations which makes no reference to the initial state. Naturally the 
resulting evolution operators contain classical 
random variables. The complexity of the stochastic equations which govern these 
classical random variables depends upon the complexity of the 
commutation relations between the operators appearing in the LSE. If the 
complexity of the commutation relations is sufficiently high then the 
stochastic 
equations governing the classical random variables become too complex to solve 
analytically. Nevertheless, even if this is the case, the form of the evolution 
operator provides information regarding the type of states produced by the LSE, 
and the problem is reduced to integrating the classical stochastic equations 
numerically. We also note that the solution to an LSE provides additional 
information to that contained in the solution to the equivalent master 
equation, because it gives the state of the system for each trajectory. For 
example, 
the variance of a system operator may be calculated for each final state (ie. 
for each trajectory), and this is referred to as the {\em conditional} variance 
as it is conditional upon the results of the measurement. The overall average 
of these variances may be then be calculated. The solution to the master 
equation allows us to calculate only the variance which is obtained by 
first averaging 
the final states over all trajectories, which is, in general, quite a 
different quantity.

In the following we use as examples LSE's corresponding to the continuous 
measurement of physical observables. A term of the form 
\begin{equation}
  \dot{\rho} = \cdots -k[O,[O,\rho]] \cdots
\label{mterm}
\end{equation}
in a quantum master equation for the evolution of a density matrix, $\rho$, 
for a quantum system $S$, describes a continuous projection measurement of 
an observable $O$ of $S$. The rate at which information is gained regarding 
the observable is determined by $k$ which is a positive constant. That a 
continuous measurement of a physical observable can be described in this 
way has been demonstrated by Barchielli and co-workers~\cite{Barchielli}, 
and also by Ueda {\em et al.\ }\cite{Ueda} using a quite different approach. 
For the theory of continuous measurement the reader is referred to these 
works and references~\cite{cm0,cm1,cm2,cm3}. We refer to this measurement 
process as a continuous {\em projection} measurement because in the absence 
of any system evolution, the sole effect of this term is to reduce the 
off-diagonal elements of the density matrix to zero in the eigenbasis of 
that observable. That is, it describes, in the long time limit, a projection 
onto one of the eigenstates of the observable under observation. If, in 
addition, the observable commutes with the Hamiltonian describing the free 
evolution of the system under observation, then the free evolution does not 
interfere with this process of projection, and the measurement is referred 
to as a continuous Quantum Non-Demolition (QND) measurement~\cite{QNDdefn,WandM}.

Before we proceed we note the following points. The LSE which is equivalent 
to the general master equation~\cite{Lindblad}
\begin{equation}
\dot{\rho} = -\frac{i}{\hbar}[H,\rho] + \sum_{n=1}^{N} (2O_n\rho O^\dagger_n
 - O^\dagger_n O_n\rho - \rho O^\dagger_n O_n),
\label{genmas}
\end{equation}
where $H$ is Hermitian and the $O_n$ are arbitrary operators, is
\begin{equation}
d|\psi\rangle = \left[ -\frac{i}{\hbar}Hdt - \sum_{n=1}^{N} (O^\dagger_n
 O_n dt - \sqrt{2}O_n dW_n(t) ) \right] |\psi\rangle ,
\label{eqlqse}
\end{equation}
where the $dW_n(t)$ are independent stochastic Wiener increments which obey 
the Ito calculus relation $dW_n(t)^2=dt$~\cite{StochMeth}.

During evolution an initialy pure quantum state remains pure, but 
changes in a random 
way determined by the values taken by the Wiener process. The state at 
time $t$, $|\psi(t)\rangle_{\mbox{\scriptsize w}}$, is not normalised, 
and the probability measure for the system to have evolved to that 
particular state at that time is given by~\cite{GandG}
\begin{equation}
 \langle \psi(t)|\psi(t)\rangle_{\mbox{\scriptsize w}} 
\mbox{d}P_{\mbox{\scriptsize w}} ,
\end{equation}
where $\mbox{d}P_{\mbox{\scriptsize w}}$ is the Wiener measure. That is, 
it is the joint probability measure for all the random variables that 
appear in the expression for $|\psi(t)\rangle_{\mbox{\scriptsize w}}$. 
It follows therefore that moments of system operators calculated 
with the equivalent master equation at time $t$ are given by the expression
\begin{equation}
\langle {\cal O}\rangle = \int\langle \psi(t)|{\cal O}
|\psi(t)\rangle_{\mbox{\scriptsize w}} \mbox{d}P_{\mbox{\scriptsize w}}.
\label{avo}
\end{equation}
where ${\cal O}$ is the system operator in question, and 
$\mbox{d}P_{\mbox{\scriptsize w}}$
represents integration over all possible values of the random 
variables. For an in-depth account of LSE's and their relationship 
to physical measurements we refer the reader to reference~\cite{GandG}.

\section{Obtaining Evolution Operators for Linear Quantum Trajectories}
\subsection{General method}
We will explicitly treat here LSE's which contain only one 
stochastic increment. However it will be clear that this treatment 
may be easily extended for multiple stochastic increments. Let us write
a general LSE with a single stochastic increment as
\begin{equation}
d|\psi\rangle = (\tilde{A} \; dt + B \; dW(t))|\psi\rangle . 
\label{lqse}
\end{equation}
In this equation $\tilde{A}$ and $B$ are arbitrary operators. 
We will see that the complexity of the evolution operator will depend 
upon the 
complexity of the commutation relations between $\tilde{A}$ and $B$.

Let us define the integral of Wiener increments over a time $\Delta t$ 
as $\Delta W(t)$. The probability density for $\Delta W(t)$ is~\cite{StochMeth}
\begin{equation}
P(\Delta W(t)) = \frac{1}{\sqrt{2\pi\Delta t}}
e^{-(\Delta W(t))^2/(2\Delta t)},
\end{equation}
so that $\langle \Delta W(t)\rangle = 0$ and $\langle 
(\Delta W(t))^2\rangle = \Delta t$.
As a first step in obtaining an evolution operator for the LSE 
in Eq.(\ref{lqse}) we rewrite it in the form
\begin{eqnarray}
|\psi(t+dt)\rangle & = & e^{(\tilde{A}-(B^2/2))dt}e^{BdW(t)}
|\psi\rangle \nonumber \\
                   & = & e^{Adt}e^{BdW(t)}|\psi\rangle 
\label{lqse2} ,
\end{eqnarray}
were we have defined $A=\tilde{A}-B^2/2$.
It is easily verified that this is correct to first order by 
expanding the exponentials to second order and using the Ito calculus 
relation $dW(t)^2=dt$.
To first order the state at time $t+\Delta t$ is therefore
\begin{equation}
|\psi(t+\Delta t)\rangle = e^{A\Delta t}e^{B\Delta W(t)}|\psi(t)\rangle 
\label{lqse3} ,
\end{equation}
so that the state at time $t$ may be written 
\begin{equation}
|\psi(t)\rangle_{\mbox{\scriptsize w}} = \lim_{\Delta t \rightarrow 0} 
\prod_{n=1}^{N} (e^{A\Delta t}e^{B\Delta W_n}) \; |\psi(0)\rangle , 
\label{lqse4}
\end{equation}
where
\begin{equation}
\Delta W_n=\int_{(n-1)\Delta t}^{n\Delta t} \!\!\!\!\!\!\!\!\!
\!\!\! dW(t) ,
\end{equation}
and $N\rightarrow \infty$ as $\Delta t \rightarrow 0$ so that 
$N\Delta t=t$ is always true. To complete the derivation of the 
evolution operator we must take the limit in Eq.(\ref{lqse4}). 
To do this we must combine the arguments of the exponentials which 
appear in the product, so that we may sum the infinitessimals. We 
will choose to do this by first repeatedly swapping the order of 
the exponentials containing the operator $A$ with those containing 
the operator $B$. The simplest case occurs when $A$ and $B$ commute 
so that the problem essentially reduces to the single variable 
case, and 
we treat this in Sec.{\ref{QNDPH}}. The simplest non-trivial case 
occurs when the commutator $[A,B]$, while non-zero, commutes with 
both $A$ and $B$, and we treat this in Sec.{\ref{pmeas}}. In the 
final part of this section  we examine a more complicated example 
in which the commutator $[A,B]$ does not commute with either $A$ or 
$B$.

\subsection{A QND measurement of photon number}\label{QNDPH}
The mathematically trivial case occurs when $A$ and $B$ commute. A 
non-trivial
physical example to which this corresponds is a QND measurement of 
the photon number of a single cavity mode. Denoting the annihilation 
operator describing the mode by $a$, the free cavity field Hamiltonian 
is given by~\cite{WandM}
\begin{equation}
  H = \hbar\omega(a^\dagger a + \frac{1}{2}),
\end{equation}
in which $\omega$ is the frequency of the cavity mode, and the 
observable to be measured is $O = a^\dagger a$. With this we have
\begin{eqnarray}
  A & = & -i\omega(a^\dagger a + \frac{1}{2}) -  2k(a^\dagger 
a)^2 , \\
  B & = & \sqrt{2k}a^\dagger a ,
\end{eqnarray}
in which $k$ is the measurement constant introduced in 
Eq.(\ref{mterm}).
As $A$ and $B$ commute the exponentials in Eq.(\ref{lqse4}) 
combine trivially and we obtain
\begin{eqnarray}
|\psi(t)\rangle_{\mbox{\scriptsize w}} & = & \lim_{\Delta t 
\rightarrow 0} e^{A N\Delta t}\exp\left[{B \sum_n \Delta W_n}\right] 
\; |\psi(0)\rangle \nonumber \\
 & = & e^{A t}e^{B W(t)} \; |\psi(0)\rangle . \label{sol1}
\end{eqnarray} 
As the Wiener process, $W(t)$, is a sum of independent Gaussian 
distributed random 
variables, $W_n$, it is naturally Gaussian distributed, the mean 
and variance of $W(t)$ being zero and $t$ respectively. In a 
particular realisation of the stochastic equation Eq.(\ref{lqse}), 
the Wiener process will have a particular value at each time 
$t$, and as we mentioned 
above, the set of all these values corresponds to the trajectory that 
is taken by that particular realisation. The fact that to obtain the 
state at time $t$ we require only the value of the Wiener process 
at that time means that we do not require all the trajectory information, 
but just a single variable associated with that trajectory. For more 
complicated cases, in which the operators do not commute, we will 
find that other variables associated with the trajectory appear in 
the evolution operator.

\begin{figure} 
\begin{center} 
\leavevmode\includegraphics[width=0.98\hsize]{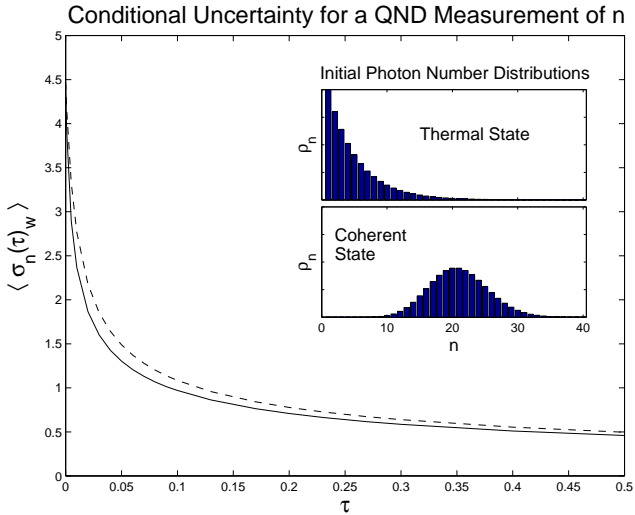}
\caption{The conditional uncertainty in photon number averaged 
over all trajectories, $\langle \sigma_n(t)_{\mbox{\scriptsize w}}\rangle$, 
is plotted here against the dimensionless scaled time, $\tau=kt$. The dotted 
line corresponds to an initial coherent state, and the solid line to an 
initial thermal state. Both initial states were chosen to have $\sigma_n^2=20$, 
giving the thermal state a mean photon number of $\langle n\rangle=4$, and the 
coherent state a mean photon number of $\langle n\rangle=20$. The photon number 
distributions for the two initial states are displayed in the inset.}
\label{nqndfig} 
\end{center} 
\end{figure}

As the situation we consider here is a QND measurement, the phase 
uncertainty introduced by the measurement of photon number does 
not feed back to affect the measurement, so that the result is simply 
to decrease continuously the uncertainty in photon number, and the 
state of the system as $t$ tends to infinity tends to a number state. 
If we denote the evolution operator derived in Eq.(\ref{sol1}) by 
$V(t)$, and start the system in an arbitrary mixed state $\rho(0)$, 
then at time $t$ the normalised state of the system may be written
\begin{equation}
   \rho(t) = \frac{V(t)\rho(0) V^\dagger(t)}{\mbox{Tr}\left\{ 
V(t)\rho(0) V^\dagger(t)\right\} }.
\end{equation}
As $V(t)$ is diagonal in the photon number basis, we only require 
the diagonal elements of the initial density matrix to calculate moments 
of the photon number operator. Denoting the diagonal elements of the 
initial density matrix by $\rho_n$, and the diagonal elements of 
$V(t)V^\dagger(t)$ by $V_n$, the variance of the photon number, for a 
given trajectory, is given by
\begin{equation}
  \sigma_{\scriptsize n}^2(t)_{\mbox{\scriptsize w}} = 
\frac{\sum_n n^2\rho_nV_n}{\sum_n \rho_nV_n} - 
\frac{\left(\sum_n n\rho_nV_n\right)^2}
     {\left(\sum_n \rho_nV_n\right)^2} .
\end{equation}
The uncertainty in our knowledge of the number of photons is the square 
root of this variance. Averaging this uncertainty over all trajectories 
therefore tells us, on average, how accurately we will have determined 
the number of photons at a later time. To calculate the value of the 
uncertainty for each trajectory, averaged over all trajectories we must 
multiply $\sigma_{\scriptsize n}(t)_{\mbox{\scriptsize w}}$ by the 
probability for each final state and average over all the final states. 
The probability measure for the final states, $\rho(t)$, is given by the 
Wiener measure multiplied by the norm of the final state, 
$\mbox{Tr}\left\{ V(t)\rho(0) V^\dagger(t)\right\}$. This probability measure 
is not in general Gaussian in $W(t)$, but a weighted sum of Gaussians, one for 
each $n$. Performing the multiplication, we obtain the average conditional 
uncertainty in photon number as
\begin{equation}
   \langle \sigma_{\scriptsize n}(t)_{\mbox{\scriptsize w}}\rangle = 
   \int \sqrt{\sum_{nm} n(m-n)\rho_n\rho_mV_nV_m} \; 
\mbox{d}P_{\mbox{\scriptsize w}} ,
\end{equation}
in which
\begin{eqnarray}
  V_n & = & e^{-4ktn^2 + 2\sqrt{2k}nW} , \\
  \mbox{d}P_{\mbox{\scriptsize w}} & = & \frac{1}{\sqrt{2\pi t}} 
e^{-W^2/(2t)}\mbox{d}W.
\end{eqnarray}
We note that 
$\langle \sigma_{\scriptsize n}(t)_{\mbox{\scriptsize w}}\rangle$ may be 
written as a function of $\tau =kt$, being the time scaled by the measurement 
constant. Hence, as we expect, increasing the measurement time has the same 
effect on 
$\langle \sigma_{\scriptsize n}(t)_{\mbox{\scriptsize w}}\rangle$ as 
increasing the measurement constant.
We evaluate 
$\langle \sigma_{\scriptsize n}(\tau)_{\mbox{\scriptsize w}}\rangle$ 
numerically for an initial thermal state, and an initial coherent state, and 
display the results in figure~\ref{nqndfig}. We have chosen the initial states 
so that they have the same uncertainty in photon number, with the result that 
the mean number of photons in each of the two states is quite different. The 
results show the decrease in uncertainty with time, which is seen to be only 
weakly dependent upon the initial state. 

\subsection{A measurement of momentum in a linear potential}\label{pmeas}
The simplest mathematically non-trivial case occurs when the commutator 
between $A$ 
and $B$, while non-zero, is such that it commutes with both $A$ and $B$. 
A physical situation to which this corresponds is a continuous measurement 
of the momentum of a particle in a linear potential. If we denote the 
position and momentum operators for the particle as $Q$ and $P$ respectively, 
then the Hamiltonian is given by
\begin{equation}
H = \frac{1}{2m}P^2 - FQ,
\end{equation}
in which $m$ is the mass of the particle and $F$ is the force on the 
particle from the linear potential. In this case we have
\begin{eqnarray}
  A & = & (\frac{-i}{2\hbar m} - 2k)P^2 + \frac{iF}{\hbar}Q , \\
  B & = & \sqrt{2k}P ,
\end{eqnarray}
in which $k$ is again the measurement constant.

Returning to Eq.(\ref{lqse4}) we see that to obtain a solution we must 
pass all the exponentials containing the operator $B$ to the right through 
the exponentials containing the operator $A$. In order to perform this 
operation we need a relation of the form
\begin{equation}
e^{B}e^{A} = e^{A}e^{C} . \label{oprel}
\end{equation}
For the present case the required relation is simply given by the 
Baker-Campbell-Hausdorff formula~\cite{BCH,BHE}
\begin{equation}
e^{B}e^{A} = e^{A}e^{B}e^{-[A,B]} .
\end{equation}
Using this relation to propagate successively all of the exponentials 
containing $B$ to the right in the product in Eq.(\ref{lqse4}) we obtain
\begin{eqnarray}
& & \prod_{n=1}^{N} (e^{A\Delta t}e^{B\Delta W_n}) = 
\exp\left[{AN\Delta t}\right]\exp\left[{B \sum_{n=1}^N\Delta W_n} 
\right] \nonumber \\ 
 & & \times \exp\left[{-[A,B]\Delta t \sum_{n=1}^N(n-1)\Delta W_n} \right] .
\end{eqnarray}
All that remains is to calculate the joint probability density for the 
random variables. The first is the Wiener process, and the second is
\begin{eqnarray}
Y(t) & = & \lim_{\Delta t \rightarrow 0}\Delta t \sum_{n=1}^N(n-1)
\Delta W_n = \int_0^t t' dW(t') . 
\end{eqnarray}
Clearly these are both Gaussian distributed with zero mean and 
all that we require is to calculate the covariances 
$\langle Y(t)^2\rangle$ and $\langle 
W(t)Y(t)\rangle$ to determine completely the joint density 
at time $t$. Using $\langle \Delta W_n \Delta W_n \rangle = 
\delta_{nm}\Delta t$ these quantities are easily obtained:
\begin{eqnarray}
\langle Y(t)^2\rangle & = & \lim_{\Delta t \rightarrow 0}\Delta t 
\sum_{n=1}^N((n-1)\Delta t)^2 \Delta t \nonumber \\ & = & 
\int_0^t t'^2 \; dt' = t^3/3 , \\
\langle W(t)Y(t)\rangle & = & \lim_{\Delta t \rightarrow 0} 
\sum_{n=1}^N((n-1)\Delta t) \Delta t \nonumber \\ & = & 
\int_0^t t' \; dt' = t^2/2 .
\end{eqnarray}
The state at time $t$, under the evolution described by the 
stochastic equation, is 
therefore
\begin{equation}
|\psi(t)\rangle_{\mbox{\scriptsize w}} = e^{At}e^{BW(t)}
e^{-[A,B]Y(t)}|\psi(0)\rangle ,
\end{equation}
where the joint probability density for $W$ and $Y$ at time $t$ 
is given by 
\begin{eqnarray}
  P_{\mbox{\scriptsize w}}(W,Y) = \left( \frac{\sqrt{12}}{2\pi t^2} \right) 
\exp\left[-\frac{2}{t}W^2 - \frac{6}{t^3} Y^2 + \frac{6}{t^2}WY \right] . 
\nonumber 
\end{eqnarray}
Note that to obtain the probability density for the final state, this must be multiplied by the norm of the state at time $t$.                                                                 

Returning to the specific case of a particle in a linear potential, we may 
now obtain results for various quantities of interest. Writing the evolution
operator in terms of the momentum and position operators we have
\begin{eqnarray}
|\psi(t)\rangle_{\mbox{\scriptsize w}} &  = &  \exp\left[{((\frac{-i}{2\hbar 
m} - 2k)P^2 + \frac{iF}{\hbar}Q)t}\right]  \nonumber \\
                                       & \times & \exp\left[\sqrt{2k}
(PW(t)+FY(t))\right] |\psi(0)\rangle .
\end{eqnarray}
Using the Zassenhaus formula~\cite{Witschel} to disentangle the 
argument of the first exponential we may rewrite this in 
the more convenient form
\begin{eqnarray}
|\psi(t)\rangle_{\mbox{\scriptsize w}}  & = & \exp\left[{ 
\frac{iF}{\hbar}Qt}\right]  \exp\left[{\eta (-P^2t - PFt^2 - 
F^2 t^3/3)}\right]  \nonumber \\
                & \times &  \exp\left[{ \sqrt{2k}(PW(t)+FY(t))}
\right]  |\psi(0)\rangle ,
\label{psit2}
\end{eqnarray}
in which $\eta = (\frac{i}{2\hbar m} + 2k)$. For those not familiar 
with the Zassenhaus formula, it is complementary to the BCH formula. 
While the BCH formula shows how to write an exponential of the sum of 
two operators as a product of exponentials of the operators and their 
commutator (or in more complicated cases repeated commutators of the 
two operators), the Zassenhaus formula shows how to write the product 
of exponentials of two operators as the exponential of the sum of the 
operators and repeated commutators.

Let us first take an arbitrary initial state, writing it in the momentum 
eigenbasis so that we have
\begin{equation}
|\psi(0)\rangle = \int_{-\infty}^{\infty} \!\!\!\!\! \Psi(p)|p\rangle  
\; \mbox{d}p  \;\; , \;\;\;  \int_{-\infty}^{\infty} \!\!\!\!\! |\Psi(p)|^2 
\; \mbox{d}p = 1 .
\label{psi0}
\end{equation}
Using Eqs.(\ref{psit2}) and (\ref{psi0}) in Eq.(\ref{avo}) to calculate the 
moments of $p$ given by first averaging over all trajectories (that is, 
the moments which would be given by the equivalent master equation) we 
readily obtain
\begin{equation}
 \langle p(t)^n\rangle = \langle (p(0)+Ft)^n\rangle .
\end{equation}
In particular, for any initial state, $|\psi(0)\rangle$, the average value 
of the momentum at time $t$, $\langle p(t)\rangle$, is simply shifted from 
the initial value by the impulse $Ft$. The variance of the momentum at time 
$t$, $\sigma_p^2(t) = \langle p(t)^2\rangle - \langle p(t)\rangle^2$, 
remains equal to its original value. That is, the uncertainty introduced into 
the position of the particle by the momentum measurement does not feed back 
into the momentum, even though the momentum does not commute with the 
Hamiltonian. This is because while the momentum determines the position 
at a later time, the converse is not true. These results for the moments 
are easily checked using the equivalent master equation.

Now let us consider the conditional variance of the momentum at time $t$
averaged over all trajectories. In the previous section we calculated the 
conditional uncertainty, being the square root of the variance, and 
averaged this over all trajectories. Here however, we will find that the 
conditional variance is independent of the trajectory taken, and depends 
only on the measurement time. This will also be true of the example which 
we will treat in the next section. In this case clearly it does not matter 
if we first average the conditional variance over the trajectories, and 
then take the square root, or if instead we average the conditional 
uncertainty, because the averaging procedure is redundant. However, in 
general the two procedures are not equivalent.  We will denote 
the conditional variance by 
$\langle\sigma_p^2(t)_{\mbox{\scriptsize w}}\rangle$. As the uncertainty in 
position does not feed back into the momentum, we expect that this variance 
should steadily decrease to zero. This is because during a trajectory our 
knowledge of the momentum steadily increases so that the distribution over 
momentum becomes increasingly narrow. To perform this calculation 
we take the initial state to be the minimum uncertainty wave-packet given by 
the ground state of a harmonic oscillator of frequency $\omega$. The average 
values of the position and momentum of the particle are both zero in this 
state and the respective variances are 
\begin{eqnarray}
 \langle Q^2\rangle & = & \frac{\hbar}{2m\omega} \equiv \sigma_Q^2(0) \;\; 
, \;\;\; \langle P^2\rangle =  \frac{m\hbar\omega}{2} \equiv \sigma_p^2(0) 
\nonumber ,
\end{eqnarray}
and in  momentum space the state may be written
\begin{equation}
 |\psi(0)\rangle = \left(\frac{1}{\pi m\hbar\omega}\right)^{\frac{1}{4}} 
\int_{-\infty}^{\infty} \!\!\!\!\! e^{-P^2/(2m\hbar\omega)} |p\rangle   
\; \mbox{d}p.
\end{equation}
The moments of momentum for each trajectory are given by
\begin{equation}
 \langle p^n\rangle_{\mbox{\scriptsize w}} = \frac{\langle\psi(t)|p^n|\psi(t)\rangle_{\mbox{\scriptsize w}}}{\langle\psi(t)|\psi(t)\rangle_{\mbox{\scriptsize w}}},
\end{equation}
and we calculate the first and second to give 
$\sigma_p^2(t)_{\mbox{\scriptsize w}} = \langle 
p(t)^2\rangle_{\mbox{\scriptsize w}} - \langle p(t)
\rangle^2_{\mbox{\scriptsize w}}$. We obtain
\begin{equation}
  \langle\sigma_p^2(t)_{\mbox{\scriptsize w}}\rangle = 
\frac{\sigma_p^2(0)}{1 + 8k\sigma_p^2(0)t } .
\end{equation}
This is independent of $W$ and $Y$ and hence independent of the 
trajectory. It is therefore unnecessary to average over the final 
states. 
Indeed $\langle\sigma_p^2(t)_{\mbox{\scriptsize w}}\rangle$ decreases 
steadily from the initial value to zero as $t\rightarrow\infty$ as we 
expect from the discussion above. This means that while the average value 
of momentum is determined by the measurement results, the error in our 
estimate of the momentum at time $t$ is not.

\subsection{A quadrature measurement with a general quadratic 
Hamiltonian}\label{quadm}
We now consider an LSE in which the 
commutator $[A,B]$ does not commute with either $A$ or $B$. As in the 
previous example, let $P$ and $Q$ be, respectively, the canonical 
momentum and position operators for a single particle so that they obey 
the canonical commutation relation $[Q,P]=i\hbar$. With this definition 
we will take $A$ and $B$ to have the following forms:
\begin{eqnarray}
A & = & \alpha P^2 + \gamma Q^2 + \xi QP + \eta P + \zeta Q , 
\label{aexp} \\
B & = & kQ+\kappa P , \label{bexp}
\end{eqnarray}
where $\alpha, \gamma, \eta, \zeta, k$ and $\kappa$ are complex numbers. 
This example applies to an optical mode of the electromagnetic field, 
including classical driving and/or classically driven subharmonic 
generation~\cite{subharm} and for which an arbitrary quadrature is 
continuously measured. It also applies to the situation of a single 
particle, which may feel a linear and/or harmonic potential, and which 
is subjected to continuous observation of an arbitrary linear combination 
of its position~\cite{atfm} and momentum.

To obtain an evolution operator for the LSE with this choice of the 
operators $A$ and $B$, we require, as before, a relation of the form 
given by Eq.(\ref{oprel}). To derive this relation we proceed in the 
following manner.

 First we may use the Baker-Campbell-Hausdorff expansion~\cite{BHE}, or 
alternatively solve the equations of motion given by $dB/d\epsilon = 
[A,B]$, to obtain an expression for $e^{\epsilon A}Be^{-\epsilon A}$. 
The result is
\begin{equation}
e^{-\epsilon A}\varepsilon Be^{\epsilon A} = \varepsilon f_1(\epsilon) 
Q +\varepsilon f_2(\epsilon) P + \varepsilon f_3(\epsilon) , 
\label{orel1}
\end{equation}
in which
\begin{eqnarray}
f_1(\epsilon) & = & \frac{1}{\lambda} (-2\kappa\gamma + k\xi) S 
+ k C, \\
f_2(\epsilon) & = & \frac{1}{\lambda} (2k\alpha - \kappa\xi) S 
+ \kappa C, \\
f_3(\epsilon) & = & \frac{1}{\lambda^2} (k\eta\xi + 2k\alpha\zeta
- \kappa\zeta\xi - 2\kappa\gamma\eta) [C-1] \nonumber \\ 
  & & + \frac{1}{\lambda} (k\eta 
+ \kappa\zeta) S .
\end{eqnarray}
In these expressions $C=\cosh(i\hbar\lambda\epsilon )$, $S=
\sinh(i\hbar\lambda\epsilon )$ and $\lambda=\sqrt{\xi^2-4\alpha\gamma}$. 
Using the relation
\begin{equation}
e^{-\epsilon A}f(\varepsilon B)e^{\epsilon A} = f(e^{-\epsilon A}
\varepsilon Be^{\epsilon A}) ,
\end{equation}
we obtain from Eq.(\ref{orel1})
\begin{equation}
e^{-\epsilon A}e^{\varepsilon B}e^{\epsilon A} = e^{\varepsilon 
f_1(\epsilon) Q +\varepsilon f_2(\epsilon) P}e^{\varepsilon f_3
(\epsilon)} .
\end{equation}
Multiplying both sides of this equation on the left by 
$e^{\epsilon A}$ we obtain a relation of the form $e^{\varepsilon B}
e^{\epsilon A}=e^{\epsilon A}e^{\varepsilon D(\epsilon)}$, as we require.

We see from the above procedure that the relation in Eq.(\ref{oprel}) 
may be obtained so long as a closed form can be found for the solution 
to the operator differential equation $dB/d\epsilon = [A,B]$. Clearly 
this is straightforward if this equation is linear in $B$, which is 
true in the example we have treated here, and is sometimes possible in 
cases in which the equations are non-linear.

In addition, for this example we also require the BCH relation in the 
form
\begin{equation}
e^{A}e^{B}=e^{A+B+\frac{1}{2}[A,B]} . \label{BCH}
\end{equation}
This is so that we can sum up in one exponential the operators that 
result from swapping $e^{\Delta W_n B}$ and $e^{n\Delta t A}$. 

Using the expressions derived above, with the replacements 
$\epsilon=n\Delta t$ and $\varepsilon=\Delta W_n$, for each $n$ from 1 
to $N$, by repeatedly swapping the exponentials containing $B$ with 
those containing $A$ as in the previous example, we obtain
\begin{eqnarray}
\lim_{\Delta t \rightarrow 0}\prod_{n=1}^{N} (e^{A\Delta t}
e^{B\Delta W_n}) & = & 
e^{At}e^{X_1(t) Q + X_2(t) P} \nonumber \\ 
 & & \times e^{X_3(t)}e^{i\hbar Z(t)} ,
\end{eqnarray}
in which the classical stochastic variables $X_i$ and $Z$, are 
given by
\begin{eqnarray}
 X_i(t) & = & \int_0^t \!\!\! f_i(t') dW(t'), \\
 Z(t) & = & \int_0^t \!\!\! f_1(t')X_2(t') dW(t') - \int_0^t \!\!
\! f_2(t')X_1(t') dW(t'), \nonumber
\end{eqnarray}
where the expressions for the $f_i$ are given above, and the 
integrals are Ito integrals. The $X_i$ are Gaussian distributed with 
zero mean, and their covariances are easily calculated as in the 
previous example:
\begin{equation}
\langle X_i(t)X_j(t)\rangle = \!\! \int_{0}^{t}\!\!\!\! f_i(t')
f_j(t')\; dt' . 
\end{equation}
In addition, the two-time correlation functions for these variables 
are also easily obtained analytically. In particular we have
\begin{equation}
\langle X_i(t)X_j(\tau)\rangle = \!\! \int_{0}^{\mbox{min}(t,\tau)}
\!\!\!\! f_i(t')f_j(t')\; dt' . 
\end{equation}
However, $Z(t)$ is not Gaussian distributed. We are not aware of an 
analytic expression for this variable, so that its probability 
density may have to 
be obtained numerically. We note in passing, however, that in some 
cases double stochastic integrals of this kind may be written explicitly 
in terms of products of Gaussian variables~\cite{StochMeth}. We note also that 
$Z$ determines only the normalisation of the final state, and not the state 
itself. The normalised state at time $t$ is therefore independent of $Z$, and 
we examine the consequences of this in appendix~\ref{apA}.

We may now write the state at time $t$ as
\begin{equation}
|\psi(t)\rangle_{\mbox{\scriptsize w}} = e^{At}e^{X_1(t)Q + X_2(t)P}
e^{X_3(t) + i\hbar Z(t)} |\psi(0)\rangle .
\label{eop3}
\end{equation}
 Hence even though values for averages 
over all trajectories may in general have to be calculated numerically,  
the evolution operator provides us with information regarding the type 
of states that will occur at time $t$. In particular, if the initial 
state is 
Gaussian in position (and therefore also Gaussian in momentum), then as 
each of the exponential operators in the above equation transform 
Gaussian states to Gaussian states, we see that the state of the system 
remains Gaussian at all times. The mean of the Gaussian in 
both position and momentum change with time in a random way determined 
by the values of the stochastic variables.

We will shortly consider a particular example: that of a harmonic oscillator 
undergoing a continuous observation of position, and use this evolution 
operator to calculate the conditional variance for the position at time 
$t$. We will take the initial state to be a coherent state, which is a 
Gaussian wave packet. This conditional variance does not depend upon the 
trajectory, but simply upon the initial state and the measurement time, 
as indeed we found to be the case for the momentum measurement in 
section~\ref{pmeas}.

Let us first show that for an initial coherent state the conditional 
variance of any linear combination of position and momentum is 
independent of the trajectory for all of the cases covered by the 
evolution operator in Eq.(\ref{eop3}). To do this we must calculate 
the effect of this evolution operator on a coherent state. Clearly 
the effect of the right-most exponential operator is at most to 
change the normalisation, which effects neither the average values 
of position and momentum, nor the respective variances. The effect 
of the next exponential, being linear in $P$ and $Q$, is calculated 
in appendix~\ref{apB}. We find that it changes the mean values of the 
position and momentum, and alters the normalisation, but the state 
remains coherent in that the position variance (and hence the momentum 
variance) is unchanged. Finally, 
the effect of the exponential quadratic in $P$ and $Q$ is calculated 
in appendix~\ref{apB}. We find that this operator modifies the position 
variance. However, as the operator does not contain any stochastic 
variables, and as the manner in which it changes the position variance 
is independent of the mean position and momentum, we obtain the result 
that the effect on the position variance, and hence the variance of 
any linear combination of position and momentum, is trajectory 
independent. 

Let us now consider a harmonic oscillator in which the position is 
continuously 
observed. This situation has been analysed by Belavkin and 
Staszewski using the equivalent non-linear equations~\cite{BS}. 
The operators $A$ and $B$ in this case are given by
\begin{eqnarray}
  A & = & \left( \frac{-i}{2\hbar m} \right)P^2 + \left( \frac{-im\omega^2}
{2\hbar}  - 2k \right)Q^2 , \\
  B & = & \sqrt{2k}Q ,
\end{eqnarray}
in which $m$ is the mass of the particle, $\omega$ is the frequency of 
the harmonic oscillation, and $k$ is the measurement constant for the 
continuous observation of position. Taking the initial state to be 
coherent, and denoting it $|\alpha\rangle$, the initial position 
wave-function is given by 
\begin{equation}
 \langle x|\alpha\rangle = \left( \frac{2s^2}{\pi} \right) ^{1/4} 
e^{-s^2x^2+2sx\alpha -\frac{1}{2}(|\alpha|^2 + \alpha^2)}, 
\end{equation}
where $s^2=m\omega/(2\hbar)$. Using the results in 
appendix~\ref{apB}, we 
find that the coefficient of $x^2$ at a later time $t$ is given by
\begin{equation}
 s'^2 = s^2 \left[ \frac{1-2l}{3-2l} \right] \left[ 1 + 
2\frac{1-2l}{1+2l} \right]
\end{equation}
where
\begin{equation}
  l = \frac{-1/2}{rz\coth(z\omega t) + (1 + ir)} ,
\end{equation}
and we have defined the parameters
\begin{equation}
  z = \sqrt{\frac{2i}{r} - 1} \;\; , \;\;\; r = \frac{m\omega^2}
{2\hbar k} .
\end{equation}
After some algebra this may be written as
\begin{equation}
 s'^2 = s^2 iz \frac{iz\tanh(z\omega t) - 1}{\tanh(z\omega t) - iz} ,
\label{deltax}
\end{equation}
in agreement with that derived by Belavkin and Staszewski.
The conditional variance for $x$ at time $t$ is given by
\begin{equation}
 \sigma_x^2(t)_{\mbox{\scriptsize w}} = 
\frac{1}{4\mbox{\small Re}[s'^2]} .
\end{equation}
As $t$ tends to infinity, Eq.(\ref{deltax}) gives a steady state 
value for the conditional variance, which is
\begin{equation}
 \sigma_x^2 = \frac{1}{4\mbox{\small Im}[z]} =  \left( \sqrt{2}s^2 
\sqrt{\sqrt{4/r^2 + 1} + 1} \right)^{-1}.
\end{equation}
The parameter $r$ is a dimensionless quantity which gives essentially 
the ratio between the frequency of the harmonic oscillator, and the rate 
of the position measurement. We may view the dynamics of the position 
variance as being the result of two competing effects. One is the action 
of the measurement which is continuously narrowing the distribution in 
position, and consequently widening the distribution in momentum. The 
other is the action of the harmonic motion, which rotates the state in 
phase space, so converting the widened momentum distribution into 
position. Depending on the relative strengths of these two processes, 
determined by the dimensionless constant $r$, a steady state is 
reached in which they balance. If the rate of the measurement is very 
fast compared to the frequency of the oscillation (corresponding to 
$r \ll 1$), then the localisation in position is much greater than it 
would be for an unmonitored oscillator, and in that case we succeed 
effectively in tracking the position of the particle. However, if the 
frequency of oscillation is much greater than the rate of localisation 
do to the measurement, then the steady state position variance remains 
essentially that of the unmonitored oscillator.

\section{Conclusion}
We have presented a method for obtaining evolution operators for various 
classes of stochastic equations describing linear quantum trajectories, 
and applied this to a number of physical examples pertaining to physical 
systems subjected to the continuous projection measurement of an observable. 
We have shown how the complexity of the stochastic equations governing the 
random variables which appear in the evolution operator depends upon the 
commutation relations between the operators appearing in the LSE. For the 
case in which both these operators commute with their commutator, 
probability densities for the random variables may be obtained 
analytically. We have also shown that in cases in which the commutation relations are more complex it is sometimes still possible to obtain an explicit evolution operator. This is possible even in cases in which the classical stochastic integrals, or equivalently the stochastic equations, governing the random variables which appear in this operator are too complex to solve analytically. 

\section*{Acknowledgments}
The authors would like to thank H.\ M.\ Wiseman for helpful comments on 
the manuscript. This work was supported in part by the U.K.\ Engineering 
and Physical Sciences Research Council, and the European Union. KJ would 
like to acknowledge support from the British Council and the New Zealand 
Vice Chancellors' Committee.

\appendix
\section{Eliminating variables which affect only the final normalisation}
\label{apA}
We found in section~\ref{quadm} that not all the random variables which 
appear in the evolution operator are Gaussian distributed. This result is 
surprising because it has been shown previously, using the non-linear 
equations, that for an initial Gaussian state, the probability density 
for the conditional mean position and momentum, and therefore for the 
final state, {\em are} Gaussian distributed for this case~\cite{solNLSE}. 
These two results may be reconciled due to the fact that the non-Gaussian 
variable in the evolution operator given in Eq.(\ref{eop3}) affects purely 
the normalisation of the final state, rather than the state itself.

Let us assume that we have an initial state $|\psi\rangle$, and an 
evolution operator which is a function of the random variables $X$ and 
$Z$ (which may in general be vector valued). We let the random variable 
$Z$ determine only the normalisation of the final state, so that the 
evolution operator may be written
\begin{equation}
   V(X,Z,t) = {\cal O}(X,t)f(Z,t) ,
\end{equation}
where ${\cal O}$ is an operator valued function, and $f$ is simply a 
complex valued function. The unnormalised state at time $t$ is then 
given by
\begin{equation}
  |\psi(t)\rangle_{\mbox{\scriptsize w}} = {\cal O}(X,t)f(Z,t)
|\psi\rangle .
\end{equation}
Clearly once we have normalised that state at time $t$, it is no-longer 
dependent upon $Z$. In particular the normalised state is given by
\begin{equation}
  |\tilde{\psi}(t)\rangle_{\mbox{\scriptsize w}} = 
\frac{ {\cal O}(X,t)|\psi\rangle_{\mbox{\scriptsize w}}}
        {\sqrt{ \langle \psi |{\cal O}^\dagger(X,t)
                              {\cal O}(X,t)
|\psi\rangle_{\mbox{\scriptsize w}} }} .
\end{equation}
The probability density for the final state is 
\begin{equation}
  P(X,Z,t) = \langle \psi(t)|\psi(t)
\rangle_{\mbox{\scriptsize w}} P_{\mbox{\scriptsize w}}(X,Z,t) ,
\end{equation}
in which $P_{\mbox{\scriptsize w}}(X,Z,t)$ is the probability 
density given by the Wiener measure 
for the variables $X$ and 
$Z$. However, seeing as the normalised state depends only upon $X$, 
we require for all calculations only the marginal probability density 
for $X$. Denoting this marginal density also by $P$, we have
\begin{equation}
  P(X,t) = \int P(X,Z,t) \; \mbox{d}Z .
\end{equation}
In certain cases the probability measure for the normalised state may 
therefore be Gaussian, even though the measure for the unnormalised 
state is not. However, as $P(X,Z,t)$ contains a factor of the 
norm of 
$|\psi(t)\rangle_{\mbox{\scriptsize w}}$, the probability measure for 
the output process will, in general, only be Gaussian if the norm is 
Gaussian in $X$. Clearly the norm is Gaussian in $X$ for initial Gaussian 
states in the case we investigate in section~\ref{quadm}.

\section{The effect on a coherent state of exponentials linear and 
quadratic in P and Q}
\label{apB}
We first calculate the effect of an operator of the form
\begin{equation}
 e^{\nu P + \mu Q}
\end{equation}
on a coherent state $|\alpha\rangle$. The coherent state is defined as 
the eigenstate of the annihilation operator $a$, such that
\begin{equation}
   a|\alpha\rangle = \alpha |\alpha\rangle ,
\end{equation}
and
\begin{equation}
   a = \sqrt{\frac{m\omega}{2\hbar}} \; x + i\sqrt{\frac{1}{2\hbar 
m\omega}} 
\; p.
\end{equation}
Here $m$ and $\omega$ are the mass and frequency of a harmonic 
oscillator 
which serves for the purposes of defining the coherent state.
 In particular we are interested in the position wave-function of 
the result. 
We therefore wish to calculate
\begin{equation}
 \langle x|\psi\rangle = \langle x|e^{\nu P + \mu Q}|\alpha\rangle ,
\end{equation}
where $|x\rangle$ is an eigenstate of the position operator $Q$ such 
that
\begin{equation}
   Q|x\rangle = x |x\rangle .
\end{equation}
Note that in general $|\psi\rangle$ will not be normalised. To 
perform this 
calculation we will need the BCH formula given in Eq.(\ref{BCH}), 
and the 
position wavefunction for a coherent state,
\begin{eqnarray}
 \langle x|\alpha\rangle & = & \left( \frac{2s^2}{\pi} \right) ^{1/4} 
e^{-s^2x^2+2sx\alpha -\frac{1}{2}(|\alpha|^2 + \alpha^2)} \nonumber \\
                         & = & \left( \frac{2s^2}{\pi} \right) ^{1/4} 
e^{-s^2x^2+2sx\alpha - \alpha_r^2 - i\alpha_r\alpha_i}
\end{eqnarray}
where
\begin{eqnarray}
 s & = & \sqrt{\frac{m\omega}{2\hbar}} \\
 \alpha & = & \alpha_r + i\alpha_i .
\end{eqnarray}
Note that this expression contains the phase factor $- i\alpha_r\alpha_i$. 
This is left out in many texts, but is essential for consistency with the 
completeness relations for the position states. We also require the inner 
product of two coherent states,
\begin{equation}
 \langle \alpha |\beta\rangle = e^{-\frac{1}{2}(|\alpha|^2 + |\beta|^2) + 
\alpha^* \beta},
\end{equation}
and the well known integral formula
\begin{equation}
  \int e^{-\alpha x^2 -\beta x} \; \mbox{d}x = \sqrt{\frac{\pi}{\alpha}} 
e^{\beta^2/(4\alpha)} \;\; , \;\; \mbox{Re}[\alpha] > 0.
\end{equation}
We proceed first by rewriting the exponential in terms of annihilation and 
creation operators, so that we have
\begin{equation}
  e^{\nu P + \mu Q} = e^{\theta a + \phi a^\dagger} = e^{\phi a^\dagger} 
e^{\theta a}e^{\theta\phi/2}
\end{equation}
in which
\begin{equation}
 \theta = \left( \nu\sqrt{\frac{\hbar}{2m\omega}} - 
i\mu\sqrt{\frac{m\hbar\omega}{2}} \right) = \phi^* .
\end{equation}
We may now use the completeness relation for the coherent states to obtain
\begin{eqnarray}
 \langle x|\psi\rangle & = & \langle x|e^{\phi a^\dagger} 
e^{\theta a}|\alpha\rangle e^{\theta\phi/2} \nonumber \\
  & = & \frac{1}{\pi} \int \!\!\!\! \int \langle x|\beta\rangle  
\langle \beta|e^{\phi a^\dagger} e^{\theta a}|\alpha\rangle 
e^{\theta\phi/2} \;\mbox{d}^2\beta \nonumber \\
 & = & \frac{1}{\pi} \int \!\!\!\! \int \langle x|\beta\rangle  
\langle \beta|\alpha \rangle \; e^{\theta \alpha + \phi \beta^* + 
\theta\phi/2} \;\mbox{d}^2\beta \nonumber \\
 & = &   \langle x|\alpha + \phi\rangle e^{\frac{1}{2}|\phi|^2+
\mbox{\scriptsize Re}[\alpha\phi^*] + \theta\alpha + \theta\phi/2}.
\end{eqnarray}
We see that the state remains coherent, although it is no longer 
normalised, and is shifted in phase space by $\phi$.

We now wish to calculate the effect of an operator of the form
\begin{equation}
 e^{\eta P^2 + \zeta Q^2 + \xi QP}
\end{equation}
on a coherent state. This time we require to calculate
\begin{equation}
 \langle x|\psi\rangle = \langle x|e^{\eta P^2 + \zeta Q^2 + 
\xi QP}|\alpha\rangle .
\end{equation}
For this calculation we will need the disentangling theorem for 
the exponential of a general quadratic form of the annihilation and 
creation operators, which is given by~\cite{mjcollett}
\begin{equation}
 e^{ua^2 + va^{\dagger 2} + wa^\dagger a} = e^{(\chi-w)/2}e^{la^{\dagger 
2}}e^{\chi a^\dagger a}e^{ma^2},
\end{equation}
in which
\begin{eqnarray}
 l & = & \frac{v}{f\coth(f) - w} \\
 \chi & = & \ln \left( \frac{f}{f\cosh(f) - w\sinh(f)} \right)   \\
 m & = & \frac{u}{f\coth(f) - w}   \\
 f & = &  \sqrt{w^2 - 4uv} .
\end{eqnarray} 
First of all rewriting the exponential containing $P$ and $Q$ as an 
exponential in the annihilation and creation operators, we have
\begin{equation}
  \langle x|e^{\eta P^2 + \zeta Q^2 + \xi QP}|\alpha\rangle = 
\langle x|e^{ua^2 + va^{\dagger 2} + wa^\dagger a + u}|\alpha\rangle
\end{equation}
in which
\begin{eqnarray}
 u & = & v^* = \left( \frac{\zeta\hbar}{2m\omega} - \frac{\eta m\hbar\omega}{2} 
- i\frac{\xi\hbar}{2} \right) \\
 w & = & \left( \frac{\zeta\hbar}{m\omega} + \eta m\hbar\omega \right) .
\end{eqnarray}
We now proceed by using the disentangling theorem, and employing the 
completeness relation for the coherent states.
\begin{eqnarray}
 \langle x|\psi\rangle & = & \langle x|e^{(w+\chi)/2}e^{la^{\dagger 2}}
e^{\chi a^\dagger a}e^{ma^2}|\alpha\rangle \nonumber \\
 & = & \frac{1}{\pi} \int \!\!\!\! \int \langle x|\beta\rangle  
\langle \beta| e^{(w+\chi)/2}e^{la^{\dagger 2}}e^{\chi a^\dagger a}
e^{ma^2}|\alpha\rangle \;\mbox{d}^2\beta \nonumber \\
 & = & \frac{1}{\pi} \int \!\!\!\! \int \langle x|\beta\rangle  
\langle \beta|\alpha e^{\chi}\rangle \; e^{l\beta^{*2}} \;\mbox{d}^2\beta 
\nonumber \\
 & &  \;\; \times \; e^{\frac{1}{2}|\alpha|^2(|e^{2\chi}|-1) + m\alpha^2} 
e^{(\chi-w)/2} .
\end{eqnarray}
Performing the integral over the real and imaginary parts of $\alpha$, 
we obtain \begin{eqnarray}
  \langle x|\psi\rangle & = & \frac{1}{\sqrt{1+2l}} \left( \frac{2s^2}
{\pi} \right) ^{\frac{1}{4}} \!\!\! e^{-\frac{1}{2}|\alpha|^2 - m\alpha^2}
 e^{(\chi-w)/2} \nonumber \\
 & & \times \exp \left\{ -s^2x^2 \left[ \frac{1-2l}{3-2l} \right] 
\left[ 1 + 2\frac{1-2l}{1+2l} \right] \right\} \nonumber \\
 & & \times \exp \left\{ 2sx\alpha e^\chi \left[ \frac{1}{3-2l} 
\right] \left[ 1 + 2\frac{1-2l}{1+2l} \right] \right\} \nonumber \\
 & & \times \exp \left\{\alpha^2e^{2\chi} \left[ \frac{1}{3-2l} 
\right] \left[ \frac{1}{2} + \frac{2}{1+2l} \right] \right\}
\end{eqnarray}
It is easily verified that this reduces to $\langle x|\alpha\rangle$ 
as required when we set $l=\chi=m=0$.

\end{document}